\documentclass[useAMS,usenatbib]{mn2e}

\usepackage{graphicx}
\usepackage{amssymb}
\usepackage{amsmath}
\usepackage{lscape}   
\usepackage{threeparttable} 
\usepackage{dcolumn}

\def\gcc{\hbox{\rm\hskip.35em  g cm}$^{-3}$}
\def\rads{\hbox{\rm\hskip.35em  rad s}$^{-1}$}
\def\radss{\hbox{\rm\hskip.35em  rad s}$^{-2}$}

\def\ergs{\hbox{\rm\hskip.35em  erg s}$^{-1}$}

\newcommand{\apj}{ApJ}
\newcommand{\apjl}{ApJ}
\newcommand{\apjs}{ApJS}

\newcommand{\mnras}{MNRAS}
\newcommand{\nat}{Nature}

\newcommand{\aap}{A{\&}A}

\newcommand{\prb}{Phys. Rev. B}
\newcommand{\prc}{Phys. Rev. C}
\newcommand{\prd}{Phys. Rev. D}
\newcommand{\prl}{Phys. Rev. Lett.}
\begin{document}

\title[Postglitch exponential relaxation of radio pulsars and magnetars]{Postglitch exponential relaxation of radio pulsars and magnetars in terms of vortex creep across flux tubes}
\author[E. G\"{u}gercino\u{g}lu]{Erbil G\"{u}gercino\u{g}lu$^1$\thanks{email: egugercinoglu@gmail.com}\thanks{email: egugercinoglu@sabanciuniv.edu}\\
$^1$Sabanc{\i} University, Faculty of Engineering and Natural Sciences, Orhanl{\i}, 34956 Istanbul, Turkey}

\maketitle
\begin{abstract}
Timing observations of rapidly rotating neutron stars revealed a great number of glitches, observed both from canonical radio pulsars and magnetars. Among them, 76 glitches have shown exponential relaxation(s) with characteristic decay times ranging from several days to a few months, followed by a more gradual recovery. Glitches displaying exponential relaxation with single or multiple decay time constants are analysed in terms of a model based on the interaction of the vortex lines with the toroidal arrangement of flux tubes in the outer core of the neutron star. Model results agree with the observed timescales in general. Thus, the glitch phenomenon can be used to deduce valuable information about neutron star structure, in particular on the interior magnetic field configuration which is unaccessible from surface observations. One immediate conclusion is that the magnetar glitch data are best explained with a much cooler core and therefore require that direct Urca type fast cooling mechanisms should be effective for magnetars.

\end{abstract}
\begin{keywords}
stars: neutron - pulsars: general - stars: magnetars - stars: magnetic fields - dense matter
\end{keywords}
\section{Introduction}
\label{intro}

Pulsars are rapidly rotating neutron stars which provide us with some of the most precise time measurer in the universe that rival the best atomic clocks. As timing observations indicate, pulsars slow down steadily as they lose rotational kinetic energy in the form of magnetic dipole radiation and particle winds \citep{lyne12}. As a consequence, once a pulsar's spin period $P$ (or frequency $\nu=1/P$) and rate of change of the spin period with time $\dot P$ (or spin-down rate $\dot\nu=-\dot P/P^{2}$) are measured its rotational evolution can be predicted with high accuracy. Nevertheless, regular spin-down of pulsars in long term is observed to be punctuated by sudden spin-ups in their rotation rate $\Omega=2\pi\nu$, i.e. glitches, occasionally. Glitch events manifest in timing data as fractional changes in the rotation rate with $\Delta\Omega/\Omega \sim10^{-11}-10^{-5}$ and are usually accompanied by jumps in the spin-down rate, $\Delta \dot\Omega /\dot \Omega \sim 10^{-4}-10^{-1}$ \citep{espinoza11,yu13,shabanova13, dib14}. Both of these changes tend to relax back to the original preglitch state  fully or partially on timescales of the order of several days to a few years. The relaxation process includes initially prompt exponential decay and then a more gradual recovery with a constant second time derivative of the angular rotation velocity, $\ddot\Omega$. Glitches are seen from pulsars of all ages but the largest ones are predominantly observed in younger pulsars with characteristic ages between $10^3$ and $10^5$ years. 

Glitch phenomena occur almost instantaneously, the best constrained 2000 Vela glitch had a rise time upper limit of only 40 seconds \citep{dodson02}. While most canonical radio pulsar glitches (for exceptions see \citet{archibald16} and references in \citet{akbal15}) are not followed by changes correlated with external torque variation \citep{espinoza11,yu13}, magnetar glitches are accompanied by radiative changes and bursting activity \citep{dib14}. This fact indicates that glitches reflect angular momentum exchange between the observed crust and interior components of the neutron star. Glitches cannot be originated from the pulsar magnetosphere as this would require drastic and irreversible changes in the neutron star structure. For instance, large glitches would require significant plasma supply to produce glitch inducing instabilities so that a great portion of the magnetosphere can be blown away due to particle wind ejecta in this course and magnetospheric changes should bring about long-lived persistent changes in pulse shapes and spectra which are not observed \citep{pines74}. However, magnetosphere may take part in magnetar glitches which show fluctuations in spindown rate, indicative of external torque variation. For the case of glitches with external torque variation, the magnetospheric processes play an important role and contribute to post-glitch relaxation along with the interior (superfluid) torques (G\"{u}gercino\u{g}lu \& Alpar, in preparation \footnote{\citet{erbil17}.}). Following the first glitch \citet{baym69} correctly interpreted that the implied rapid angular momentum exchange and significantly slow postglitch recovery provides strong evidence for the existence of a interior superfluid component weakly coupled to the neutron star crust and the magnetosphere, since a star composed of normal matter should relax faster. 

The origin of pulsar glitches remains controversial. Several mechanisms related to superfluid vortex dynamics were proposed to explain initiating a glitch, including crustquake induced unpinning \citep{alpar96}, avalanches from high density vortex traps \citep{cheng88}, coupling of vortex oscillations to lattice phonons \citep{epstein92}, extra pinning barrier for vortices sustained by interaction of crustal magnetic field with flux tubes in the outer core \citep{sedrakian99}, self-organised criticality and vortex proximity effects \citep{warszawski12}. According to the pinning-unpinning model of \citet{anderson75}, the superfluid in the inner crust with quantized vortex lines acts as angular momentum reservoir driving the glitches.  In the case of the array of vortex lines pin on the crystal lattice, they move rigidly with the crust and thus slow down, while the angular velocity of the superfluid does not change since it is solely determined by the number of vortices which remained fixed in this course. When the difference between the crustal and the superfluid angular velocities exceeds a critical value, vortices are liberated and consequently this leads to the transfer of their angular momentum to the crust, i.e. a glitch, rather rapidly. Between the glitches vortex lines slowly migrate outwards among adjacent pinning sites via thermal activation. Post-glitch relaxation occurs through reestablishment of this thermal creep conditions for the vortices depinned during a glitch \citep{alpar84a}. Vortex line pinning to and creep against atomic nuclei in the crustal lattice picture has been confirmed by recent realistic numerical computations \citep{seveso16,haskell16}.

Short glitch rise times and long recovery duration provides us with a unique opportunity to investigate a number of physical processes, including the crust-core coupling \citep{abney96,newton15}, redistribution of excess angular momentum within discrete internal layers \citep{howitt16}, spin-up in neutral and charged superfluids \citep{easson79,sourie17} and constraining the properties of the bulk matter \citep{melatos10} and the equation of state \citep{datta93, link99}. Furthermore, unhealed rotational parameters after successive glitches compared to the preglitch state may give rise to significant modifications in the dipole magnetic field estimates since $B_{\rm d} \propto (P\dot P)^{1/2}$ \citep{lin04}. In this paper by analysing the glitches displaying exponential relaxation  it is shown that glitch phenomenon can be used to deduce valuable information about neutron star internal structure, in particular on the interior magnetic field configuration. To this end, a model incorporating glitch induced offset $\Delta\Omega_{\rm c}$ and vortex line-flux tube creep process in the outer core of the neutron star is considered and its predictions are compared with the existing glitch data. Corresponding toroidal field region in the outer core as a whole decouples from the rest of the star during both at the time of glitch and post-glitch relaxation stage.

This paper is organised as follows. In Section \ref{model} we present the basics of the model describing the dynamic interaction between the vortex lines and the flux tubes and its response to  a glitch. In Section \ref{application} we confront the model predictions with the existing data on radio pulsar and magnetar glitches which showed exponential recovery. Finally, Section \ref{cad} presents a discussion of the results obtained in this paper.  
\section{The Model: Vortex Creep Across Toroidal Flux Tubes}
\label{model}

The vortex creep model \citep{alpar84a} explains both glitches and postglitch
recovery in terms of moments of inertia and response times of the
neutron superfluid permeating the lattice nuclei in the neutron star crust. The superfluid core of the star, which comprises the most of the moment of inertia, is coupled to the crust on rather short timescales \citep{alpar84b} and thus effectively included in the observed spindown of the normal matter crust and magnetosphere. In the absence of glitches, the crustal superfluid and rest of the star slow down at the same rate with a lag $\omega=\Omega_{\rm s}-\Omega_{\rm c} > 0$ between the superfluid and the crustal angular velocities. The external torque due to magnetic dipole radiation maintains the required bias that drives an average vortex current radially outward from the rotation axis through consecutive pinning and unpinning by thermal activation. By this ``vortex creep" superfluid couples to the normal matter and it allows the superfluid to spin down. If $\omega$ reaches a critical value $\omega_{\rm cr}$, pinning forces can no longer sustain the lag and a sudden discharge of the pinned vortices takes place. As a result, the superfluid rotation rate decreases by $\delta\Omega_{\rm s}$ and by angular momentum conservation the crust rotation rate increases by $\Delta\Omega_{\rm c}$, so that the lag changes by $\delta\omega=\delta\Omega_{\rm s}+\Delta\Omega_{\rm c}$ at the time of the glitch. This glitch induced reduction in $\omega$ temporally stops the creep and post-glitch relaxation occurs as a process of establishment of pre-glitch creep conditions again. 

The observed spindown rate $\dot\Omega_{\rm c}$ after the glitches exhibits several distinct components with different moments of inertia and relaxation modes. Depending on the ratio of the pinning energy $E_{\rm p}$ to the internal temperature $T$ and whether the system is close to the critical conditions for unpinning threshold, vortex creep can operate in linear or nonlinear regimes \citep{alpar89}. In the linear regime, the steady state lag $\omega_{\infty}$ is much smaller than $\omega_{\rm cr}$ and contributes an exponentially relaxing term with moment of inertia $I_{\rm l}$ to the postglitch response:
\begin{align}
\Delta\dot{\Omega}_{\rm c}(t)=-\frac{I_{\rm l}}{I}\frac{\delta\omega}{\tau_{\rm l}}{\rm e}^{-t/\tau_{\rm l}},
\label{lcreep}
\end{align}
with a relaxation time,
\begin{equation}
\label{taulin}
\tau_{\rm l} \equiv \frac{kT }{E_{\rm p}} \frac{R \omega_{\rm cr}}{4 \Omega_{\rm s} v_{0}} \exp \left( \frac{E_{\rm p}}{kT} \right),
\end{equation}
where the distance $R$ of the vortex lines from
the rotational axis is approximately equal to the neutron star radius $R_*$ and $v_{0}\approx 10^{7}$ cm/s is microscopic vortex velocity around nuclei \citep{alpar84a,erbil16}. In such a region no glitch induced vortex motion takes place and therefore $\delta\omega = \Delta\Omega_{c}$. As can be seen from equation (\ref{taulin}) $\tau_{\rm l}$ has a very strong dependence on the local pinning energy and internal temperature. By contrast, within a nonlinear creep regime $\omega_{\infty} \cong \omega_{\rm cr}$ and its contribution with moment of inertia $I_{\rm nl}$ to the postglitch response is
\begin{equation}
\Delta\dot\Omega_{\rm c}(t)=-\frac{I_{\rm nl}}{I}\vert\dot{\Omega}\vert \left[1-\frac{1}{1+({\rm e}^{t_0/\tau_{\rm nl}}-1){\rm e}^{-t/\tau_{\rm nl}}}\right],
\label{ncreep}
\end{equation}
with a nonlinear creep relaxation time
\begin{equation}
\tau_{\rm nl}\equiv \frac{kT}{E_{\rm p}}\frac{\omega_{\rm cr}}{\vert\dot{\Omega}\vert}.
\label{taun}
\end{equation}
Nonlinear creep regions are responsible for glitches through vortex unpinning avalanches and creep restarts after a waiting time $t_{0}=\delta\omega/\vert\dot{\Omega}\vert\simeq\delta\Omega_{\rm s}/\vert\dot{\Omega}\vert$. 

In recent years the vortex creep model faced with a new theoretical constraint. Dripped superfluid neutrons in the inner crust are in Bloch states of the crust lattice. Bragg scattering of free superfluid neutrons from crystal nuclei results in a somewhat reduction of the superfluid flow. This can be taken into account as neutrons gained effective mass $m_{\rm n}^{*}$ that is larger than their bare mass $m_{\rm n}$ \citep{chamel12}. This ``mass entrainment" effect leaves only a fraction of the crustal superfluid to effectively store and exchange angular momentum with the normal matter \citep{chamel06,andersson12,chamel13,delsate16}. Therefore, the actual amount of superfluid reservoir available for glitches is $\sim(m_{\rm n}/m_{\rm n}^{*})I_{\rm s}\delta\Omega_{\rm s}$. As a consequence, total moments of the inertia of the creep regions participating in a glitch must be larger than that of estimated from the fractional change in the spin-down rate by a factor of mass enhancement factor: $I_{\rm creep}/I \sim (m_{\rm n}^{*}/m_{\rm n})\Delta\dot\Omega/\dot\Omega$. The required moments of inertia in components of the star with pinning/creep can then exceed the moment of inertia of the crustal superfluid alone for reasonable neutron star equations of state \citep{andersson12,chamel13, delsate16}. This suggests the involvement of the core superfluid in glitches and postglitch relaxation. \citet{erbil14,erbil16} have given the basis of how core superfluid could contribute to both glitches and postglitch relaxation.

In the core, protons are expected to form a type II superconductor with magnetic field is quantized into flux tubes. If present at all, type I superconductivity with an admixture of field free superconducting regions and magnetic normal matter domains  exists near the centre of the star \citep{jones06}. When the core protons underwent transition into type II superconductivity, Meissner currents in the crust-core interface played a boundary condition for the coupled evolution of the magnetic field within the normal matter crust and superconducting core  which is unknown currently \citep{jones04}. \citet{spruit99} theoretically investigated the conditions under which magnetic fields inside stratified stars remained stable and found that the toroidal $B_{\phi}$ to poloidal $B_{\rm p}$ field ratio should satisfy 
\begin{equation}
\frac{B^{2}_{\phi}}{B_{\rm p}}< \frac{N r^{2}\rho^{1/2}}{l_{\rm h}},
\label{criteria}
\end{equation}
where $N$ is the buoyancy frequency of the stratified medium, $\rho$ is density, $l_{\rm h}$ is the horizontal
length scale of the perturbations which can be as large as the
stellar radius $R_*$, and $r$ is the cylindrical radial coordinate. For typical neutron star parameters \citet{erbil14} give the estimate
\begin{equation}
B_{\phi}\simeq 10^{14}\left(\frac{B_{\rm p}}{10^{12} \mbox{G}}\right)^{1/2} \mbox{G}.
\label{Btor}
\end{equation}
\citet{braithwaite09} conducted numerical simulations for stable magnetic field configurations in main sequence and neutron stars by using the criterion given by equation (\ref{criteria}) and thus equation (\ref{Btor}) is expected to be in reasonable agreement with actual strength of the toroidal field component in canonical radio pulsars. As for magnetars, the prescription given by (\ref{Btor}) presumably does not valid, see the discussions at the end of this section and also in section \ref{refmag}.

The bulk of the core proton superconductor-neutron superfluid region is likely to carry a poloidal array of flux tubes and a toroidal arrangement of flux tubes is expected to reside inside a equatorial belt surrounding the poloidal field component \citep{braithwaite09, gourgouliatos13,lander14}. The poloidal field strength is maximum at the stellar centre, while toroidal field component attains its largest value in the outer regions, at $r \gtrsim 0.5 R_*$. In the case of the poloidal arrangement of flux tubes, displacement of a flux tube with respect to a vortex line does not change the pinning energy when the corresponding structures are parallel to each other and vortex can find easy directions to pass through the flux tubes mesh \citep{konenkov01,sidery09}. This will make the effect of pinning and creep in the core dependent on the angle between the rotation and magnetic axes, which is highly variable among different pulsars. A toroidal arrangement of flux tubes, by contrast, provides a topologically unavoidable site for vortex pinning and creep, and can have similar conditions to those of the crustal lattice \citep{sidery09,erbil14}.

For a conservative neutron star magnetic field configuration the toroidal field is maximum at $r \sim 0.8R_*$, confined within an equatorial belt of radial extension $\sim 0.1R_*$ \citep{lander12,lander14}. The moment of inertia fraction controlled by vortex lines passing through the corresponding toroidal field region is estimated to be $I_{\rm tor}/I\gtrsim 5\times 10^{-2}$ \citep{erbil14,kantor14}. Depending on the radial extent of the toroidal field within the outer core, the moment of inertia of the associated region can be  larger and accommodates the extra moment of inertia required by the mass entrainment effect \citep{erbil14}.

For a laminar superfluid flow the toroidal field region has no appropriate structures to provide  vortex traps (that is small scale and high vortex density regions under extreme conditions close to unpinning avalanche) that initiate glitches. This means there is no vortex unpinning and no glitch associated vortex motion occurs inside the toroidal field region. The creep rate is therefore only affected by the offset to the angular velocity of the crust, $\Delta\Omega_{\rm c}$, alone. In this case we obtain  $t_{0}=\frac{\Delta\Omega_{\rm c}}{|\dot\Omega|}\ll \tau_{\rm tor}$. Then expanding equation (\ref{ncreep}) in $t_{0}/\tau_{\rm tor} \ll 1$ yields \citep{erbil14}
\begin{align}
\Delta\dot{\Omega}_{c}(t)=-\vert\dot{\Omega}\vert\frac{I_{\rm tor}}{I}\frac{t_{0}}{\tau_{\rm tor}}{\rm e}^{-t/\tau_{\rm tor}},
\label{dnumod}
\end{align}
with the toroidal field region's relaxation time is determined from equation (\ref{taun}) as 
\begin{align}
\tau_{\rm tor}&\simeq60\left(\frac{\vert\dot{\Omega}\vert}{10^{-10}\mbox{
\radss}}\right)^{-1}\left(\frac{T}{10^{8}\mbox{
K}}\right)\left(\frac{R}{10^{6}\mbox{ cm}}\right)^{-1}
x_{\rm p}^{1/2}\times\nonumber\\&\left(\frac{m_{\rm p}^*}{m_{\rm p}}\right)^{-1/2}\left(\frac{\rho}{10^{14}\mbox{\gcc}}\right)^{-1/2}\left(\frac{B_{\phi}}{10^{14}\mbox{ G}}\right)^{1/2}\mbox{days,}
\label{tau}
\end{align}
where $m_{\rm p}^*(m_{\rm p})$ is effective (bare) mass of protons, $x_{\rm p}$ is proton fraction of the neutron star core, and $R$ is the radius of the location of the toroidal field region. This response of the nonlinear vortex creep against toroidal flux tubes to a glitch induced offset is of the same form as the linear creep response of inner crust superfluid,
equation (\ref{lcreep}), but the nonlinear relaxation time is now given by equation (\ref{tau}), rather than equation (\ref{taulin}).

Note that as argued by \citet{erbil16} superfluid instabilities related to turbulence and r-modes may lead to development of high density vortex regions that may trigger an avalanche responsible for the largest observed glitches with $\Delta\Omega_{\rm c}/\Omega_{\rm c} \gtrsim 10^{-5}$. Although being an interesting possibility, this is not considered in the present paper and left as a further study.
 
Vortex creep across toroidal flux tubes also affects the thermal evolution of the neutron stars. Superfluid friction with normal matter becomes a dominant heating mechanism, in particular for older neutron stars. Steady state heat generation rate due to vortex creep across toroidal flux tubes array is given by 
\begin{equation}
\dot E \simeq10^{33}\left(\frac{I_{\rm tor}/I}{0.1}\right)\left(\frac{\omega_{\rm cr}}{0.1\mbox{\rads}}\right)\left(\frac{\vert\dot{\Omega}\vert}{10^{-10}\mbox{
rad/s$^{2}$}}\right) \mbox{\ergs}.
\label{heating}
\end{equation}
Note that the above heating rate is limited to the toroidal field region and does not represent the total volume average. Such an additional heating will become important especially in old systems where original internal thermal content is radiated away. In old millisecond pulsars due to reduced spin-down rate and increased thermal coupling through equation (\ref{heating}) creep relaxation timescale (\ref{tau}) becomes longer as $\tau_{\rm tor}\propto T\left|\dot\Omega\right|^{-1}$.

In some works absolute or perfect pinning of vortex lines to flux tubes was assumed \citep{link03,glampedakis11}. However, perfect pinning conditions do not realise in the neutron star core due to
\begin{enumerate}
\item{thermal fluctuations as a result of nonzero temperature and radial bias arising from the external torque induced spin-down \citep{alpar84a, sidery09}},
\item{relative motion between the vortex lines and the flux tubes \citep{ruderman98}},
\item{reduced pinning as a result of flux tube flexibility \citep{erbil16}}.
\end{enumerate}

Based on an idea put forward by \citet{easson79}, \citet{glampedakis15} proposed that closed field line region where toroidal flux resides rotates faster than the rest of the star and will be pushed into the crust in order to ensure energy minimisation, thereby diminishing the toroidal field. However, \citet{easson79} considered a non-superconducting quantum plasma for the neutron star core in his calculations and it is a well known fact that proton superconductor with embedded flux tubes achieve corotation with rest of the star at the expense of a tiny (London) magnetic field $b_{\rm L}=(2m_{\rm p}c/e)\Omega \cong2\times10^{-2}(\Omega/100 \mbox{\rads})$ G \citep{sauls89}. Also stable stratification of neutron star matter resists magnetic flux transport between adjacent layers characterised with different chemical composition \citep{reisenegger09}.  

By using a quasi-linear relation between the surface poloidal field and the interior magnetic field, it has been argued that inside the magnetars flux tubes are so closely packed that superconductivity is destroyed \citep{sedrakian15}. However, simple physical arguments show that since toroidal component surrounds poloidal field from the outer in the neutron star core, toroidal flux tubes are strongly twisted and impart some fraction of their magnetic energy to the poloidal component as the surface field is increased \citep{spruit09}. Thus, toroidal field component weakens as compared to the poloidal component and it seems that toroidal field inside magnetars may reach values as low as one hundredth of the surface dipole field as recent simulations indicate \citep{fujisawa14}. A secular evidence for interior toroidal field component of magnitude $B_{\phi}\lesssim 10^{15}$ G for magnetars comes from the neutrino emission rates \citep{suwa14}. More exotic types of superconductivity may arise or superconductivity may break down deep inside the neutron star core where toroidal field region terminates \citep{alford08}. 

Long term magnetic field evolution determines the presence and extent of the toroidal field region within the neutron star core. Following the formation of the neutron star the mixed magnetic field configuration with poloidal and toroidal components are inherited from progenitor star \citep{braithwaite09} and after the transition into superconducting state this field form is frozen into the neutron star plasma due to the enhanced electrical conductivity \citep{jones06}. Flux tubes move out as a result of diffusion processes, forces acting on them and secular interaction with vortex lines. Detailed calculations show that interior magnetic field does not decay appreciably until the crustal field is dissipated, i.e. about for $10^{6-7}$ years \citep{jahan-miri00, elfritz16}. As for field evolution in connection with the motion of vortex lines, \citet{erbil16} have shown that during creep process force exerted by a vortex line to flux tubes during successive encounters is negligible compared to buoyancy and other forces. Therefore, in glitching pulsars for which ages change in the range $10^{3}-10^{6}$ years toroidal field region remains in the outer core and will be stationary with respect to the crustal frame.    
 
So, the existence of the toroidal field region with vortex lines' creeping across flux tubes is a well established fact both for canonical radio pulsars and magnetars. Analysis of the vortex velocity around flux tubes tangle in the outer core shows that the vortex creep against flux tubes is always in the nonlinear regime \citep{erbil16} and its response to a glitch is exponential relaxation with a timescale given by equation (\ref{tau}) \citep{erbil14}. As will be shown in this paper the toroidal field region within which vortices creep through not only complements the extra moment of inertia necessity brought by mass entrainment effect but also fits both radio pulsar and magnetar post-glitch relaxation observations by giving exponential decay. 
\section{Application to the Post-Glitch Relaxations of the Radio Pulsars and the Magnetars}
\label{application}

The time-dependence of the pulsar frequency after a glitch is generally well described by the following function \citep{wang00}:
\begin{equation}
\nu \left(t\right)=\nu_0 \left(t\right)+ \Delta \nu_{\rm g}\left[1-Q\left(1-{\rm e}^{-t/\tau_{\rm d}}\right)\right]+\Delta\dot\nu_{\rm p}t,
\label{nuobs}
\end{equation}
where $\nu_0 $ is extrapolated frequency from the pre-glitch state, $\Delta \nu_{\rm g}=\Delta \nu_{\rm d}+\Delta \nu_{\rm p}$ is the total glitch magnitude in which $\Delta \nu_{\rm d}$ decays exponentially with a time constant $\tau_{\rm d}$ and $\Delta \nu_{\rm p}$ is permanent or long term increase in pulse frequency while $\Delta\dot\nu_{\rm p}$ is permanent change in the spin-down rate. Quite important quantity, the healing parameter $Q=\Delta \nu_{\rm d}/\Delta\nu_{\rm g}$ quantifies the degree of relaxation towards pre-glitch conditions. In the two component glitch model \citep{baym69}, the healing parameter is identified as ratio of the moment of inertia of the superfluid component to that of the charged normal matter of the neutron star, $Q=I_{\rm s}/I_{\rm c}$. If there exists more than a single exponentially decaying component, their contributions to $Q$ are evaluated separately. $Q$ values have a bimodal distribution. When all the glitching pulsars are considered, it is found that there is a trend indicating high $Q$ values for younger pulsars and low $Q$ values for older ones \citep{yu13}. 

The change in the spindown rate at the time of a glitch is given by the time derivative of the equation (\ref{nuobs}) 
\begin{eqnarray}
\label{dnuobs}
\Delta\dot\nu \left(t\right)&=&\Delta\dot\nu_{\rm d}\left({\rm e}^{-t/\tau_{\rm d}}\right)+ \Delta\dot\nu_{\rm p}\\ &=&\frac{-Q\Delta \nu_{\rm g}}{\tau_{\rm d}}\left({\rm e}^{-t/\tau_{\rm d}}\right)+\Delta\dot\nu_{\rm p} \nonumber.
\end{eqnarray}
Here $\Delta\dot\nu_{\rm d}=-\Delta\nu_{\rm d}/\tau_{\rm d}$.
Upon comparing  the fit function (\ref{dnuobs}) with vortex creep across flux tubes model prediction equations (\ref{dnumod}) and (\ref{tau}) one arrives at three possibilities:
\begin{enumerate}
\item If $\tau_{\rm tor} \approx \tau_{\rm d}$, then $I_{\rm tor}/I \sim Q$.
\item If $\tau_{\rm tor} \gg \tau_{\rm d}$ but $Q \ll 1$, then the relaxation of the toroidal field region is not completed yet and one can only say $I_{\rm tor}/I \lesssim 1-Q$.
\item If $\tau_{\rm tor} \ll \tau_{\rm d}$ while glitch date uncertainty is large, then the prompt relaxation of the toroidal flux region is over and should simply be missed from the observations. 
\end{enumerate}

In order to compare the toroidal field region's relaxation timescale given by equation (\ref{tau}) with the existing glitch data, a cooling law specifying the internal temperature is needed. For the modified Urca process \citet{yakovlev11} derived the following analytical fit assuming a nonsuperfluid neutron star model:
\begin{equation}
T_{\rm in}= 1.96\times10^{8}~{\rm K}~{\rm e}^{-\Phi(r)}\left(1-x\right)\left(1+0.12R_{6}^{2}\right)\left(\frac{10^{4}{\rm yrs}}{t_{\rm age}}\right)^{1/6},
\label{Tmod}
\end{equation}
where $\Phi(r)$ is the metric function quantifying the gravitational redshift, $x=2GM_*/c^{2}R_*$ is the compactness of the star, $M_*$ is the mass of the neutron star, $R_{6}=R_*/10^{6}$ cm and $t_{\rm age}=\nu/2\left|\dot\nu\right|$ is characteristic (spindown) age of the pulsar. Throughout the computations a neutron star model with a radius $R_*=12$ km and mass $M_*=1.6M_{\odot}$ is adopted. Then equation (\ref{Tmod}) assumes the form
\begin{equation}
T_{\rm in}= 1.78\times10^{8}~{\rm K}~\left(\frac{10^{4}{\rm yrs}}{t_{\rm age}}\right)^{1/6}.
\label{murca}
\end{equation}
Superfluidity affects the efficiency of neutrino emission, reduces the heat capacity of the star and thus brings about regulations in the cooling law through changing the coefficient in equations (\ref{Tmod}) and (\ref{murca}) while $t_{\rm age}^{-1/6}$ behaviour is signature of the modified Urca process and does not depend on the superfluid properties. On the other hand, \citet{aguilera08} have shown that the available surface temperature measurements of neutron stars can be fitted by a model including the superfluid effects and the Joule heating with the following simple cooling law:
\begin{equation}
T_{{\rm s},6}^{4} \simeq CB_{14}^2,
\label{Tapm}
\end{equation}
where $T_{{\rm s},6}$ is the surface temperature in units of $10^{6}$ K, $B_{14}$ is the surface dipolar magnetic field in units of $10^{14}$ G and $C\simeq10$ is a constant that depends on the thickness of the crust, ohmic decay timescale, and ratio of the internal field to the surface dipole component. Translating the surface temperature in equation (\ref{Tapm}) into internal temperature through the prescription of \citet{gudmundsson83} and fitting the Vela pulsar parameters gives  \citep{glampedakis09}
\begin{equation}
T_{\rm in}= 0.72\times10^{8}~{\rm K}~\left(\frac{10^{4}{\rm yrs}}{t_{\rm age}}\right)^{1/6}.
\label{agu}
\end{equation}
 
As for the effective mass of protons, \citet{alpar84b} have shown that the mass currents of two types of the superfluids in the neutron star core, the neutron superfluid and the proton superconductor, interact with each other and this results in a effective mass for protons slightly different from their bare mass. Effective to bare mass ratio is given by \citep{alpar84b}
\begin{equation}
\left(\frac{m_{\rm p}^{*}}{m_{\rm p}}\right)=\frac{\rho_{\rm p}}{\rho_{\rm s}^{\rm pp}},
\end{equation}
where $\rho_{\rm p}=\rho_{\rm s}^{\rm pp}+\rho_{\rm s}^{\rm pn}$ is the superconducting protons' density, $\rho_{\rm s}^{\rm pp}$ and $\rho_{\rm s}^{\rm pn}$ are the density of bare protons and protons entrained by neutrons, respectively. \citet{borumand96} have obtained analytically $\rho_{\rm s}^{\rm pp}\approx 2m_{\rm p}n_{\rm p}$ and $\rho_{\rm s}^{\rm pn}\approx -0.04m_{\rm n}n_{\rm n}$ within a factor of two. Here  $m_{\rm n} (m_{\rm p})$  and $n_{\rm n} (n_{\rm p})$ are mass and number densities of neutrons (protons). Since $m_{\rm p}^{*}/m_{\rm p}<1$, proton mass current is inversely directed with the neutron velocity field. \citet{chamel08} has shown that effective to bare mass ratio for \citet{douchin01} and \citet{akmal98} equations of state changes in the range $m_{\rm p}^{*}/m_{\rm p} \approx 0.9-0.4$ throughout the entire core depending upon the nucleon-nucleon interactions which in turn depends on the density.

In the literature detailed analysis of 41 pulsars which exhibited 76 glitches with exponential decay have been published. Of these, 60 glitches observed with one exponentially decaying component while 14 glitches are found to have two exponential decay time constants. One of the most studied pulsars, the Vela pulsar has shown three distinct exponential transients in its 2 giant glitches. Glitch properties for radio pulsars are displayed in Table (\ref{tab:exp}) while the ones for magnetars are shown in Table (\ref{magnetar}).    

\subsection{Results for Radio Pulsars}
\label{radio}
The model highlighted in the preceding section was successfully applied to the all giant glitches of the Vela pulsar \citep{erbil14,akbal16} and 2007 peculiar glitch of PSR J1119--6127 \citep{akbal15} which were evaluated in terms of the vortex creep model.  For the Vela pulsar $\tau_{\rm tor}\approx 35$ days and $I_{\rm tor}/I=(0.32-1.24)\times 10^{-2}$ values were obtained while for the peculiar glitch of PSR J1119--6127 $\tau_{\rm tor}\approx 50$ days and $I_{\rm tor}/I=1.74\times 10^{-1}$ were found. Now the model presented in Section \ref{model} is applied to the whole sample of available data for radio pulsars which exhibited at least one exponentially decaying component in their glitches. To this purpose, equation (\ref{tau}) is evaluated for three models for which parameters used are listed below. The aim is to show how the toroidal field region's relaxation time changes with the employed equation of state and temperature profile adopted as well as the radial extent of the toroidal field region. Magnetic field strength is determined from the scaling of equation (\ref{Btor}).     

{\em Model 1--} Employed equation of state is \citet{akmal98}. Temperature evolution is chosen as given by equation (\ref{murca}). Microscopic parameters related to effective masses are taken from \citet{chamel08}. Toroidal flux tubes are considered to located in a torus which is extending to $R=0.6R_*$. Then, $\rho \approx 8\times 10^{14}$\gcc, $x_{\rm p}\approx 0.1$ and $m_{\rm p}^{*}/m_{\rm p}\approx 0.4$.

{\em Model 2--} Employed equation of state is \citet{lattimer91}. Temperature evolution is chosen as given by equation (\ref{murca}).  Microscopic parameters related to effective masses are determined from the parametrization of \citet{borumand96}. The response of the toroidal field region is considered to be dominated by a site close to the crust-core interface, i.e. $R=0.9R_*$. Then, $\rho \approx 1.5\times 10^{14}$\gcc, $x_{\rm p}\approx 0.04$ and $m_{\rm p}^{*}/m_{\rm p}\approx 0.55$.  

{\em Model 3--} Employed equation of state is \citet{douchin01}. Temperature evolution is chosen as given by equation (\ref{agu}). Microscopic parameters related to effective masses are taken from \citet{chamel08}. The toroidal field is considered to reach its maximum intensity at a region with $R=0.8R_*$. Then, $\rho \approx 2.5\times 10^{14}$\gcc, $x_{\rm p}\approx 0.05$ and $m_{\rm p}^{*}/m_{\rm p}\approx 0.8$.

Results for the radio pulsars are shown in Table (\ref{tab:exp}). As can be seen from Table (\ref{tab:exp}) equation (\ref{tau}) accounts for the majority of radio pulsar glitch observations quite well. For the group of pulsars PSR B0525$+$21, PSR J1141$-$6545, PSR B1727$-$47, PSR B1809$-$173, PSR B1838$-$04 and PSR J1853$+$0545 the toroidal field region's relaxation timescale given by equation (\ref{tau}) is substantially larger than the observed exponential decay times. But since for the corresponding group $Q\sim0.01-0.1$ and this can reflect the fact that for these glitches observations were interrupted before the relaxation towards preglitch conditions is eventually completed. This is consistent with the item (ii) in Section \ref{application}. For four exceptions, PSR J0205$+$6449 PSR J1112$-$6103, PSR J1420$-$6048 and PSR J1846$-$0258, the toroidal field region's relaxation time is considerably shorter than the observed decay timescales.  As can be clearly seen from Table (\ref{tab:exp}) the corresponding timescales, 4, 12, 7 and 6 days, respectively are shorter than the glitch date uncertainty and thus can simply be missed from observations as indicated by item (iii). Note that a putative glitch with such a short recovery timescale accompanying observed outburst may be responsible for the abrupt change in the braking index of PSR J1846--0258 reported by \citet{archibald15}.
\begin{landscape}
\begin{table}
\caption{Radio pulsar glitches displaying postglitch exponential decay are confronted with the model. First column encodes pulsars' name, second column indicates their characteristic ages, third and fourth column gives surface dipole (at equator) and interior toroidal field strengths, respectively. While the parameters belonging to glitches, i.e. their date, magnitudes, jumps in the spin down rate and healing parameters are displayed in  fifth, sixth, seventh and eight columns, respectively. The observed exponential decay timescales are shown in the ninth column while toroidal relaxation timescale for three different models are given in remaining tenth, eleventh and twelfth columns. Glitch data are taken from \citet{manchester05} and ATNF Glitch Table (http:www.atnf.csiro.au/people/pulsar/psrcat/glitchTbl.html).}
\label{tab:exp}
\begin{center}
\begin{threeparttable}
{\tiny
\begin{tabular}{lcccccccccccD{.}{.}{1}D{.}{.}{0}D{.}{.}{4}l}
\hline\\
\multicolumn{1}{c}{Pulsar} & \multicolumn{1}{c}{Age} & \multicolumn{1}{c}{$B_{\rm d}$} & \multicolumn{1}{c}{$B_{\phi}$} & \multicolumn{1}{c}{Glitch Date} & $\Delta\nu_{\rm g}/\nu$ & $\Delta\dot{\nu}_{\rm g}/\dot{\nu}$ & $Q$ & $\tau_{\rm d}$ (d) & $\tau_{\rm tor}$ (d) & $\tau_{\rm tor}$ (d) & $\tau_{\rm tor}$ (d) \\
& \multicolumn{1}{c}{($10^{4}$\,yr)} & \multicolumn{1}{c}{($10^{12}$\,G)} & \multicolumn{1}{c}{($10^{14}$\,G)} & \multicolumn{1}{c}{(MJD)} & ($10^{-9}$) & ($10^{-3}$) & (obs) &  (obs) & (Model 1) & (Model 2) & (Model 3) \\ 

\hline\\
J0205$+$6449 & 0.54 & 3.61 & 1.9 & 52920(144) & 5400(1800) & 52(1) & 0.77(11) & 288(8)  & 14 & 12 & 4 \\\\
B0355$+$54 &  56 & 0.84 & 0.92 & 46497(8) & 4368(2) & 96(17) & 0.00117(4) & 160(8) & 1100 & 936 & 270 \\\\
B0525$+$21 & 148 & 12.4 & 3.52 & 42057(14) & 1.2(2) & 2(2) & 0.6(2) & 140(80) & 115551 & 98325 & 28400 \\
& & & & 52280(4) & 1.6(2) & 1.1(1) & 0.44(5) & 650(50) & 115551 & 98325 & 28400 \\\\
B0531$+$21 &  0.12 & 3.79 & 1.95 & 40494 & 4.0(3) & 0.116(19) & 0.6(1) & 18.7(1.6) & 2 & 1.8 & 0.5 \\
& & & & 42447.5 & 43.8(7) & 2.15(19) & 0.8(1) & 18(2) & 2 & 1.8 & 0.5 \\
& & & & & & & 0.536(12) & 97(4) & 2 & 1.8 & 0.5 \\
& & & & 46664.4 & 4.1(1) & 2.5(2) & 1.00(4) & 9.3(2) & 2 & 1.8 & 0.5 \\
& & & & &  & & 0.89(9) & 123(40) & 2 & 1.8 & 0.5 \\
& & & & 47767.4 & 85.1(4) & 4.5(5) & 0.894(6) & 18(2) & 2 & 1.8 & 0.5 \\
& & & & & & & 0.827(5) & 265(5) & 2 & 1.8 & 0.5 \\
& & & & 48947.0(2) & 4.2(2) & 0.32(3) & 0.87(18) & 2.0(4) & 2 & 1.8 & 0.5 \\
& & & & 50020.6(3) & 2.1(1) & 0.20(1) & 0.8$^{+0.3}_{-0.2}$ & 3.2$^{+7.3}_{-2.2}$ & 2 & 1.8 & 0.5 \\
& & & & 50259.93$^{+0.25}_{-0.01}$ & 31.9(1) & 1.73(3) & 0.680(10) & 10.3(1.5) & 2 & 1.8 & 0.5 \\
& & & & 50459.15(5) & 6.1(4) & 1.1(1) & 0.87(6) & 3.0$^{+0.5}_{-0.1}$ & 2 & 1.8 & 0.5 \\
& & & & 50812.9$^{+0.3}_{-1.5}$ & 6.2(2) & 0.62(4) & 0.9(3) & 2.9(1.8) & 2 & 1.8 & 0.5 \\
& & & & 51452.3$^{+1.2}_{-1.6}$ & 6.8(2) & 0.7(1) & 0.8(2) & 3.4(5) & 2 & 1.8 & 0.5 \\\\
J0631$+$1036 &  4.36 & 5.55 & 2.36 & 52852.0(2) & 19.1(6) & 3.1(6) & 0.62(5) & 120(20) & 385 & 327 & 95 \\
& & & & 54632.41(14) & 44(1) & 4(2) & 0.13(2) & 40(15) & 385 & 327 & 95 \\\\
B0833$-$45 & 1.13 & 3.38 & 1.84 & 40280(4) & 2338(9) & 10.1(3) & 0.001980(18) & 10(1) & 35 & 30 & 9 \\
& & & & & & & 0.01782(5) & 120(6) & 35 & 30 & 9 \\
& & & & 41192(8) & 2047(30) & 14.8(2) & 0.00158(2) & 4(1) & 35 & 30 & 9 \\
& & & & & & & 0.01311(9) & 94(5) & 35 & 30 & 9 \\
& & & & 41312(4) & 12(2) & 1.9(2) & 0.1612(15) & 10.0(5) & 35 & 30 & 9 \\
& & & & 42683(3) & 1987(8) & 11(1) & 0.000435(5) & 4.0(4) & 35 & 30 & 9 \\
& & & & & & & 0.003534(16) & 35(2) & 35 & 30 & 9 \\
& & & & 43693(12) & 3063(65) & 18.3(2) & 0.00242(2) & 6.0(6) & 35 & 30 & 9 \\
& & & & & & & 0.01134(2) & 75(3) & 35 & 30 & 9 \\
& & & & 44888.4(4) & 1138(9) & 8.43(6) & 0.000813(8) & 6.0(6) & 35 & 30 & 9 \\
& & & & & & & 0.00190(4) & 14(2) & 35 & 30 & 9 \\
& & & & 45192.1(5) & 2051(3) & 23.1(3) & 0.002483(7) & 3.0(6) & 35 & 30 & 9 \\
& & & & & & & 0.00550(8) & 21.5(2.0) & 35 & 30 & 9 \\
& & & & 46259(2) & 1346(5) & 6.16(3) & 0.0037(5) & 6.5(5) & 35 & 30 & 9 \\
& & & & & & & 0.1541(6) & 332(10) & 35 & 30 & 9 \\
& & & & 47519.80360(8) & 1805.2(8) & 77(6) & 0.005385(10) & 4.62(2) & 35 & 30 & 9 \\
& & & & & & & 0.1684(4) & 351(1) & 35 & 30 & 9 \\
& & & & 50369.345(2) & 2110(17) & 5.95(3) & 0.030(4) & 186(12) & 35 & 30 & 9 \\
& & & & 51559.3190(5) & 3152(2) & 495(37) & 0.0088(6) & 0.53(3) & 35 & 30 & 9 \\
& & & & & & & 0.00547(6) & 3.29(3) & 35 & 30 & 9 \\
& & & & & & & 0.006691(7) & 19.07(2) & 35 & 30 & 9 \\
& & & & 53193.09 & 2088 & 737 & 0.009 & 0.23 & 35 & 30 & 9 \\ 
& & & & & & & 0.0056 & 2.1& 35 & 30 & 9 \\ 
& & & & & & & 0.0068 & 26.14 & 35 & 30 & 9 \\ 
& & & & 53959.93 & 2620 & 230(40) & 0.0119(6) & 73(8) & 35 & 30 & 9 \\\\
B1046$-$58 &  2.03 & 3.49 & 1.87 & 49034(9) & 2995(7) & 1.0(4) & 0.026(6) & 160(43) & 78 & 66 & 19 \\ 
& & & & 50788(3) & 771(2) & 4.62(6) & 0.008(3) & 60(20) & 78 & 66 & 19 \\\\
J1052$-$5954 & 14.32 & 1.92 & 1.39 & 54495(10) & 495(3) & 86(14)  & 0.067(4) & 46(8) & 499 & 424 & 123 \\\\
J1112$-$6103 & 3.27 & 1.45 & 1.2 & 53337(30) & 1202(20) & 7(2) & 0.022(2) & 302(146) & 49 & 42 & 12 \\\\
J1119$-$6127 &  0.16 & 41 & 6.4 & 53290 & 330(40) & 6.1(4) & 0.84(3) & 41(2) & 57 & 49 & 14 \\
& & & & 54240 & 1670(30) & 180(40) & 0.81(4) & 15.7(3) & 57 & 49 & 14 \\
& & & & & & & 0.214(7) & 186(3) & 57 & 49 & 14 \\\\
J1123$-$6259 & 81.9 & 1.21 & 1.1 & 49705.87(1) & 749.12(12) & 1.0(4) & 0.0026(1) & 840(100) & 2854 & 2429 & 702 \\\\
J1141$-$6545 & 144.9 & 1.32 & 1.15 & 54277(20) & 589.0(6) & 5.0(9) & 0.0040(7) & 495(140) & 6811 & 5796 & 1674 \\\\
B1259$-$63 & 33.21 & 0.34 & 0.58 & 50690.7(7) & 3.20(5) & 2.5(1) & 0.36(8) & 82(18) & 172 & 	146 & 42 \\\\

\end{tabular}}
\end{threeparttable}
\end{center}
\end{table}
\end{landscape}

\addtocounter{table}{-1}

\begin{landscape}
\begin{table}
\caption{--- {\it continued}}
\begin{center}
\begin{threeparttable}
{\tiny
\begin{tabular}{lcccccccccccD{.}{.}{1}D{.}{.}{0}D{.}{.}{4}l}
\hline\\
\multicolumn{1}{c}{Pulsar} & \multicolumn{1}{c}{Age} & \multicolumn{1}{c}{$B_{\rm d}$} & \multicolumn{1}{c}{$B_{\phi}$} & \multicolumn{1}{c}{Glitch Date} & $\Delta\nu_{\rm g}/\nu$ & $\Delta\dot{\nu}_{\rm g}/\dot{\nu}$ & $Q$ & $\tau_{\rm d}$ (d) & $\tau_{\rm tor}$ (d) & $\tau_{\rm tor}$ (d) & $\tau_{\rm tor}$ (d) \\
& \multicolumn{1}{c}{($10^{4}$\,yr)} & \multicolumn{1}{c}{($10^{12}$\,G)} & \multicolumn{1}{c}{($10^{14}$\,G)} & \multicolumn{1}{c}{(MJD)} & ($10^{-9}$) & ($10^{-3}$) & (obs) &  (obs) & (Model 1) & (Model 2) & (Model 3) \\ 

\hline\\
J1301$-$6305 & 1.1 & 7.1 & 2.67 & 51923(23) & 4630(2) & 8.6(4) & 0.0049(3) & 58(6) & 83 & 71 & 20 \\\\
B1338$-$62 & 1.21 & 7.08 & 2.66 & 48645(10) & 993(2) & 0.7(5) & 0.016(2) & 69(8) & 94 & 80 & 23 \\
& & & & 50683(13) & 703(4) & 1.2(3) & 0.0112(19) & 24(9) & 94 & 80 & 23 \\\\
J1412$-$6145 & 5.06 & 5.64 & 2.38 & 51868(10) & 7253.0(7) & 17.5 (8) & 0.00263(8) & 59(4) & 479 & 408 & 118 \\\\
J1420$-$6048 & 1.3 & 2.41 & 1.55 & 52754(16) & 2019(10) & 6.6(8) & 0.008(4) & 99(29) & 27 & 23 & 7 \\\\
J1522$-$5735 &  5.18 & 1.81 & 1.35 & 55250 & -11.4(6) & -1.2(13) & 1.4(2) & 27(5) & 119 & 101 & 29 \\\\
J1531$-$5610 & 9.71 & 1.09 & 1.04 & 51731(51) & 2637(2) & 25(4) & 0.007(3) & 76(16) & 146 & 124 & 36 \\\\
J1702$-$4310 & 1.70 & 7.43 & 2.73 & 53943(169) & 4810(27) & 17(4) & 0.023(6) & 96(16) & 158 & 134 & 39 \\\\
B1706$-$44 & 1.75 & 3.12 & 1.77 & 48775(15) & 2057(2) & 4.0(1) & 0.01748(8) & 122(3) & 55 & 47 & 14 \\
& & & & 52716(57) & 2872(7) & 8.0(7) & 0.0129(12) & 155(29) & 55 & 47 & 14 \\
& & & & 54711(22) & 2743.9(4) & 8.41(8) & 0.00849(7) & 85(2) & 55 & 47 & 14 \\\\
B1727$-$33 & 2.6 & 3.48 & 1.87 & 47990(20) & 3070(10) & 9.7(7) & 0.0077(5) & 110(8) & 108 & 92 & 27 \\
& & & & 52107(19) & 3202(1) & 5.9(1) & 0.0102(9) & 99(23) & 108 & 92 & 27 \\\\
B1727$-$47 & 8.04 & 11.79 & 3.43 & 52472.70(2) & 126.4(3) & 3.4(2) & 0.073(7) & 210(37) & 2229 & 1896 & 548 \\\\
B1737$-$30 & 2.06 & 17 & 4.13 & 50936.803(4) & 1445.5(3) & 2.6(8) & 0.0016(5) & 9(5) & 575 & 490 & 141 \\
& & & & 52347.66(6) & 152(2) & 0.1(7) & 0.103(9) & 50 & 575 & 490 & 141 \\
& & & & 53036(13) & 1853.6(14) & 3.0(2) & 0.0302(6) & 100 & 575 & 490 & 141 \\\\
B1757$-$24 & 1,55 & 4,04 & 2,01 & 49476(6) & 1990.1(9) & 5.6(3) & 0.0050(19) & 42(14) & 65 & 55 & 16 \\
& & & & 52055(7) & 3755.8(4) & 6.8(1) & 0.024(5) & 208(25) & 65 & 55 & 16 \\ 
& & & & 54661(2) & 3101(1) & 9.3(1) & 0.0064(9) & 25(4) & 65 & 55 & 16 \\\\
B1758$-$23 & 5.83 & 6.93 & 2.63 & 53309(18) & 494(1) & 0.19(3) & 0.009(2) & 1000(100) & 749 & 637 & 184 \\\\
B1800$-$21 & 1.58 & 4.29 & 2.07 & 48245(20) & 4073(16) & 9.1(2) & 0.0137(3) & 154(3) & 72 & 61 & 18 \\
& & & & 50777(4) & 3184(1) & 8.0(2) & 0.0094(11) & 12(2) & 72 & 61 & 18 \\
& & & & & & & 0.0030(17) & 69(13) & 72 & 61 & 18 \\
& & & & 53429(1) & 3929.3(4) & 10.6(1) & 0.00630(16) & 133(11) & 72 & 61 & 18 \\\\
J1809$-$1917 & 5.13 & 1.47 & 1.21 & 53251(2) & 1625.1(3) & 7.8(3) & 0.00602(9) & 126(7) & 91 & 77 & 22 \\\\
B1809$-$173 & 100 & 4.85 & 2.2 & 53105(2) & 14.8(6) & 3.6(5) & 0.27(2) & 800(100) & 21216 & 18054 & 5215 \\\\
B1823$-$13 & 2.14 & 2.79 & 1.67 & 53737(1) & 3581(1) & 9.6(4) & 0.0066(3) & 80(9) & 63 & 54 & 16 \\\\
B1830$-$08 & 14.74 & 0.9 & 0.95 & 48041(20) & 1865.9(4) & 1.8(5) & 0.0009(2) & 200(40) & 199 & 170 & 49 \\\\
B1838$-$04 & 46.13 & 1.1 & 1.05 & 53408(21) & 578.8(1) & 1.4(6) & 0.00014(20) & 80(20) & 1187 & 1010 & 292 \\\\
J1846$-$0258 & 0.07 & 48.68 & 6.98 & 53883.0(3.0) & 4000(1300) & 4.1(2) & 8.7(2.5) & 127(5) & 25 & 21 & 6 \\\\
J1853$+$0545  & 327 & 0.28 & 0.53 & 53450(2) & 1.46(8) & 3.5(7) & 0.22(5) & 250(3) & 2928 & 2492 & 720 \\\\
J1906$+$0722 & 4.92 & 2.02 & 1.42 & 55063(6) & 4538(14) &8.87(10) & 0.0089(2) & 221(12) & 128 & 109 & 32 \\\\
B2334$+$61 & 4.06 & 9.91 & 3.15 & 53615(6) & 20579.4(12) & 156(4) & 0.0046(7) & 21.4(5) & 721 & 613 & 177 \\
& & & & & & & 0.0029(1) & 147(2) & 721 & 613 & 177 \\\\
\hline\\
\end{tabular}}
\end{threeparttable}
\end{center}
\end{table}
\end{landscape} 

\subsection{Results for Magnetars}
\label{refmag}
Evaluation of the magnetar data requires a closer look at their specific features about magnetic field and internal temperature. As a result of magnetic field decay \citep{dall'osso12}, magnetars have much smaller spin down rates \citep{olausen14} \footnote{http://www.physics.mcgill.ca/pulsar/magnetar/main.html} and larger surface temperatures \citep{thompson96} compared to the canonical radio pulsars. Then, with the physical parameters chosen as in the previous subsection \ref{radio}, equation (\ref{tau}) would imply relaxation times of the order of several thousands of days. However, magnetar case requires a more careful modelling of magnetic field and temperature evolution inside magnetars. As mentioned earlier in section \ref{model}, due to magnetic energy and helicity transfer from the toroidal component to the poloidal component \citep{spruit09}, in magnetars toroidal field component may become as small as $B_{\phi}\lesssim 0.01B_{\rm p}$ compared to somewhat lager dipole field \citep{fujisawa14} and equation (\ref{Btor}) does not apply for this case. Given that for magnetar field strengths Hall drift and ambipolar diffusion driven magnetic field evolution becomes effective \citep{goldreich92}, it is not an unreasonable approximation to take $B_{\phi}\approx 0.01B_{\rm p}$.

Even though magnetars have higher inferred surface temperatures compared to radio pulsars, they can harbour a cooler core. All neutron star surface temperature observations, including the members of Low Mass X-Ray Binaries (LMXBs) seems to be explained, at least quantitatively, by referring to more powerful direct Urca cooling inside neutron stars' core if a certain central mass density threshold is exceeded \citep{bezgonov15a,bezgonov15b}. Magnetic field decay is not a likely agent to affect the thermal content of the core for magnetars. It would be fair to assume that the magnetic field decay in the core of magnetars do not have enough energy to put the core out of the equilibrium owing to higher thermal conductivity and enhanced neutrino emission \citep{beloborodov16}. Due to the isothermality considerations magnetic field assisted heating in the magnetar cores therefore can be safely neglected \citep{kaminker06,kaminker09}. Then, the thermal state of the magnetars with high inferred surface temperatures can be explained by placing the unspecified heating source (presumably due to magnetic field decay) phenomenologically inside the neutron star crust or on the surface while keeping the core much cooler \citep{kaminker14,beloborodov16}. The surface then does not feel the thermal state of the core, and the direct Urca process can operate inside the core with desired thermal equilibrium. So, magnetars can have colder cores with the direct Urca process as the dominant cooling agent if they are heated in the outer crust via a plausible mechanism. A fit for the internal temperature in the case of very powerful direct Urca process is (P.~S. Shternin, private communication)
\begin{equation}
T_{\rm int} = 6.7\times 10^{6}{\rm e}^{-\Phi({\rm r})}(1-x)(1+0.04 R_{6}^{2}) \left(\frac{t_{\rm age}}{10^{3} \mbox{yrs}}\right)^{-1/4} \mbox{ K},
\label{Tdurca}
\end{equation} 
where, the parameters have the same meanings as in equation (\ref{Tmod}). A quantitatively similar expression can be obtained by simple analytical parametrization of specific heat and neutrino emissivity \citep{page06,shternin15}:
\begin{equation}
T_{\rm int} \approx 5\times 10^{6}\left(\frac{10^{4}\mbox{ yrs}}{t_{\rm age}}\right)^{1/4} \mbox{ K}.
\label{durca}
\end{equation}
For \citet{akmal98} equation of state onset of the direct Urca process corresponds to a neutron star mass of $M_{\rm D}\gtrsim 1.7 M_{\odot}$ \citep{gusakov05}. As recent population synthesis studies indicate \citep{ferrario06}, progenitor stars of magnetars may be more massive than the canonical radio pulsars. Massive progenitor scenario for magnetars is further supported by dynamo activation for these stellar class \citep{obergaulinger14}. Also, some theoretical works relates ultrastrong magnetic fields of magnetars to a transition into a magnetized core realised in a sufficiently massive neutron star \citep{bhattacharya07}. For a $1.8 M_{\odot}$ mass neutron star \citet{akmal98} equation of state gives a stellar radius of $R_*=11.3$ km and if the toroidal field strength reaches its maximum at $R=0.8R_*$, then one obtains $\rho \approx 6\times 10^{14}$\gcc, $x_{\rm p}\approx 0.075$ and $m_{\rm p}^{*}/m_{\rm p}\approx 0.5$.
\begin{table*}
\caption{Magnetar glitches displaying postglitch exponential decay are confronted with the model. For each sources  $B_{\phi}\sim0.01 B_{\rm p}$ is assumed. Glitch data are taken from \citet{manchester05} and ATNF Glitch Table (http:www.atnf.csiro.au/people/pulsar/psrcat/glitchTbl.html).}
\label{magnetar}
\begin{center}{\tiny
\begin{tabular}{lccccccccc}
\hline\hline\\
\multicolumn{1}{c}{Magnetar} & \multicolumn{1}{c}{Age} & \multicolumn{1}{c}{$B_{\rm d}$} & \multicolumn{1}{c}{Glitch Date} & $\Delta\nu_{\rm g}/\nu$ & $\Delta\dot{\nu}_{\rm g}/\dot{\nu}$ & $Q$ & $\tau_{\rm d}$ (d) & $\tau_{\rm tor}$ (d) & $\tau_{\rm tor}$ (d) \\
& \multicolumn{1}{c}{($10^{4}$\,yr)} & \multicolumn{1}{c}{($10^{12}$\,G)} & \multicolumn{1}{c}{(MJD)} & ($10^{-9}$) & ($10^{-3}$) & & (observation) & (Modified Urca) & (Direct Urca) \\ 
\hline\\
4U 0142$+$61 & 6.8 & 134 & 53809 & 1630(350) & 5100(1100) & 1.1(3) & 17.0(1.7) & 381 & 23 \\\\
1RXS J1708$-$4009 & 0.9 & 468 & 52014.77 & 4210(330) & 546(62) & 0.97(11) & 50(4) & 166 & 12 \\\\
SGR J1822-1606 & 44 & 51 & 55756 & 230(10) & $-$ & 1.0 & 40(6) & 1079 & 55 \\\\
1E 1841$-$045 & 0.46 & 703 & 5246.400448 & 15170(711) & 848(76) & 0.63(5) & 43(3) & 125 & 10 \\\\
1E 2259+586 & 23 & 59 & 52443.13(9) & 4240(110) & -22(3) & 0.185(10) & 15.9(6) & 556 & 30 \\\\
\hline\\
\label{magnetar}
\end{tabular}}
\end{center}
\end{table*}

Results for magnetars are shown in Table (\ref{magnetar}). For completeness, equation (\ref{tau}) is evaluated for \citet{akmal98} equation of state parameters cited above with two different cooling models: modified Urca process with equation (\ref{agu}) and powerful direct Urca process with equation (\ref{durca}). An inspection of Table (\ref{magnetar}) reveals that the magnetar data can be best explained by a neutron star model with a core cooling via direct Urca process.
Note that two magnetars in Table (\ref{magnetar}), namely 1RXS J1708$-$4009 and 1E 1841$-$045 with ages $t_{\rm age} \lesssim 10^{4}$ years, are younger than the other magnetars in the sample and have shorter relaxation times $\sim10$ days accordingly. In these younger magnetars it may happen $B_{\phi}\simeq 0.1B_{\rm p}$ as $B_{\phi}/B_{\rm p}$ ratio does not decrease appreciably at early ages in connection with the long term magnetic field evolution \citep{gourgouliatos16}. Then, for 1RXS J1708$-$4009 and 1E 1841$-$045 the toroidal field region's relaxation time becomes 38 days and 30 days, respectively. After all that the agreement between the model predictions and the observed decay timescales improves.
\section{Discussion and Conclusions}
\label{cad}

In this work, the response of vortex creep against toroidal flux tubes to a glitch is confronted with the existing glitch observations in the literature. The corresponding region's response to a glitch is exponential relaxation and depends on the pulsar's macroscopic traits like spindown rate, toroidal component of the magnetic field, internal temperature as well as microscopic properties determined by employed equation of state like density, proton's  fraction and effective mass \citep{erbil14}. There exists 41 pulsars exhibited in total 76 glitches with exponential decay. Upon employing two different cooling models and using relevant microphysical parameters from three equations of state, three relaxation timescales are obtained. Then, an examination on whether these timescales fit the observed decay timescales is realised. It is understood that the toroidal field region's relaxation gives consistent results with the observed timescales. Although the entire parameter space is not investigated, with one single expression, equation (\ref{tau}), one is able to fit glitch data both from canonical pulsars and magnetars satisfactorily in line with the items considered in section \ref{application} which is a strong indicator of the applicability of the model presented in this paper. This fact alone shows that postglitch exponential relaxation is a powerful tool to probe into neutron star internal structure. One immediate conclusion is that magnetar observations are better fitted by a model with a much cooler interior, i.e. the dominant cooling mechanism is direct Urca process. Furthermore, glitch observations can be used to constrain the radial extent of the toroidal field region inside neutron stars (through $I_{\rm tor}/I$) which remains hidden to pulsar surface observations.  

With the contribution to glitches by vortex pinning and creep in the neutron star core, toroidal field region provides the extra moment of inertia required by mass entrainment effect \citep{erbil14,erbil16}. In older pulsars, relaxation times calculated by equation (\ref{tau}) become substantially longer \citep{erbil14}. In this case glitches would resemble step like permanent increments with no significant relaxation. Such a behaviour is indeed observed from the whole glitch data sample \citep{espinoza11,yu13}. 

The toroidal field region's relaxation time given by equation (\ref{tau}) have some advantages compared to the linear creep relaxation time given by equation (\ref{taulin}). First of all, equation (\ref{tau}) does not involve the uncertainties of the $E_{\rm p}$ estimate and $\omega_{\rm cr}$ while equation (\ref{taulin}) have strong dependence on these somewhat less constrained parameters. Therefore, microphysical variables of equation (\ref{tau}) are more reliably constrained upon employing a given equation of state. Secondly, due to the very sensitive exponential dependence on $E_{\rm p}$ and $T$, in principle by using equation (\ref{taulin}) it is possible to derive desired long decay timescales without obtaining stringent constraints about the neutron star physical structure. By contrast, equation (\ref{tau}) with adopting a magnetic field configuration and temperature profile gives an independent way to infer properties about equation of state of the neutron star matter. Nonetheless, the presence of linear creep regions' relaxation from some crustal superfluid regions in the timing data cannot be excluded \citep{erbil16}.  

In the literature glitch magnitude was used as a promising way to constrain the crustal moment of inertia and in turn the equation of state of the neutron star \citep{datta93,link99,andersson12,delsate16} but post-glitch relaxation is not considered for this purpose before. The novelty of the present paper is that the post-glitch exponential decay timescales can be used to deduce equation of state microphysical parameters and to determine the location of the crust-core interface.  

Contrary to the dipole or poloidal field, the toroidal field component is always confined within the crust or outer core of a neutron star and therefore it is rather difficult to obtain stringent constraints about the exact magnetic field configuration. Neutron star surface cooling observations \citep{geppert06}, deformation induced oscillations \citep{lander09} and precession \citep{wasserman03}, quasi-periodic oscillations \citep{gabler13} and magnetar bursts \citep{perna11} provide indirect restrictions on the toroidal field strength and location. As shown in this paper for the first time, glitch observations in turn give a novel opportunity to infer toroidal field characteristics pertaining to superfluid/superconducting outer core, within the vortex creep model.    

Vortex creep across toroidal flux tubes model has found a compelling support from a glitch in an old pulsar. Recently, \citet{lyne17} reported that $\sim10^{6}$ years old PSR J0611+1436 has undergone a Vela-like giant glitch with a quasi-exponential recovery timescale of about 12 years. Such a long decay time was never observed before from pulsars of any type and age. According to the model presented in this paper, as a pulsar ages response of the toroidal field region to a glitch is represented by a much longer exponential relaxation time and equation (\ref{tau}) with Model 1 in section \ref{radio} gives 9.4 years for PSR J0611+1436 parameters which fits approximately the observations. Moreover, \citet{pintore16} have analysed the data of transient magnetar XTE J1810$-$197 and shown that timing variability in quiescence can be accounted for an anti-glitch with an exponential decay timescale $\tau_{\rm d}=51 \pm 21$ days. Equation (\ref{tau}) with direct Urca cooling model in Section \ref{refmag} gives $\tau_{\rm tor}=45$ days and able to explain observations. 
\section*{Acknowledgements}

This work is supported by the Scientific and Technological Research Council of Turkey
(T\"{U}B\.{I}TAK) under the grant 113F354. Preliminary results of this paper were presented at NewCompStar 2015 Annual Meeting held in Budapest, Hungary which was supported by COST action MP1304. I am grateful to M.~Ali Alpar for fruitful discussions on this and related topics over the years. I wish to express my special thanks to Peter S.~Shternin for useful remarks on the cooling of neutron stars. I acknowledge the referee for insightful comments that lead to clarification of some points.



\end{document}